\documentclass[fleqn,twoside]{article}
\usepackage{espcrc2}

\usepackage{graphicx}
\usepackage[figuresright]{rotating}

\newcommand{\AmS}{{\protect\the\textfont2
  A\kern-.1667em\lower.5ex\hbox{M}\kern-.125emS}}

\title{Cosmography by GRBs :\\
Gamma Ray Bursts as possible distance indicators}

\author{S. Capozziello\address{Dipartimento Scienze Fisiche, Universit\'a di Napoli Federico II\\
        INFN Sez. di Napoli, Compl. Univ. M. S. Angelo, Ed. N, via Cinthia, 80126, Napoli, Italy}
        and
        L. Izzo\address{ICRANet and ICRA, Piazzale della Repubblica 10, 65122 Pescara, Italy\\
        Dip. di Fisica, Universit\'a di Roma “La Sapienza”, Piazzale Aldo Moro 5, 00185 Roma, Italy}}

\begin{document}

\begin{abstract}
A new method to constrain the cosmological equation
of state is proposed by using combined samples of gamma-ray bursts
(GRBs) and supernovae (SNeIa). The Chevallier-Polarski-Linder
parameterization is adopted for the equation of state in order to
find out  a realistic approach to achieve the
deceleration/acceleration transition phase of dark energy
models. As results, we find that GRBs, calibrated by SNeIa, could
be, at least,  good distance indicators capable of discriminating
cosmological models with respect to $\Lambda$CDM at high redshift.
\vspace{1pc}
\end{abstract}

\maketitle

\section{INTRODUCTION}

The acceleration of the universe is one of the most important scientific discoveries of the recent years, \cite{Riess}.
It led to the hypothesis of the existence of a force, called \emph{dark energy}, as responsible for this acceleration.
The observational evidence for an accelerating universe came from the analysis of the luminosity distances of Type Ia Supernovae, obtained by the well-known relationship between the luminosity and the light curve shape, \cite{Phillips}.
So one of the main research lines in observational cosmology is to reconstruct the evolution of the universe not only from a dynamical point of view, but mainly from a thermal one.

On the other hand the limits imposed by the instruments devoted to the detection of Supernovae Ia do not allow us to go beyond a certain redshift value, fixed by the redshift of the most distant supernova yet seen, $z_t = 1.7$.
Then in order to measure distances beyond this value of redshift, we need indicators observed at higher redshift than $z_t$.
One of the possible solutions to this problem is the use of Gamma Ray Bursts (GRB).
GRBs are the most powerful explosions in the universe, since they originated from the formation of a Black Hole.
Moreover they are observed at considerable distances, of the order of the redshift $z = 8-10$, so that there are several efforts to frame them into the standard of the cosmological distance ladder.

In the literature there are several detailed models which give account for the GRB formation and emission mechanisms, but none of them is intrinsically capable of connecting all the GRB observable quantities, see e.g. \cite{Meszaros}.
For this reason GRBs cannot be considered as standard candles.
However there exists several observational correlations among some photometric and spectroscopic properties of GRBs which features them to be used as distance indicators.

In this work we consider two empirical correlations between GRB's observable quantities in order to build a GRB Hubble diagram, after a calibration of these correlations with the Supernovae Ia data.
This GRB Hubble diagram represents the starting point for our analysis, in which we take into account the possibility that GRBs may be an extension of the Supernovae Ia at high redshifts.
Later on the analysis of this new sample is made in order to test the Dark Energy models in the context of the $\Lambda$CDM concordance model.

\section{THE GRB HUBBLE DIAGRAM}

In recent years, a great number of space missions have been devoted to detailed and continuous observation of the GRBs.
In particular, among the many results obtained, there were found some relations between spectroscopic and photometric quantities of GRBs themselves to the point that, using these relations, it is possible to trace back a very accurate estimate of the total energy emitted by these sources.
Nevertheless, up to now, there is no theoretical model that fully explain these relations so for this reason GRBs cannot be considered as standard candles in a proper sense.
For a detailed review of the observational features see \cite{Schaefer}.

In this work we are taking into account the existing 3-parameter relations.
These relations put the better constraints on the data giving less scatter between the theoretical relation and the experimental data.
The first relation is the Liang-Zhang relation, \cite{LZ}, which connects the isotropic energy released in the burst $E_{iso}$ with the GRB peak energy, $E_p$, and with the break-time of the X-afterglow light curve.
\begin{equation}\label{eq:noLZ}
 \log{E_{iso}}=a + b_1 \log{\frac{E_p (1+z)}{300keV}} + b_2 \log{\frac{t_b}{(1+z)1day}}
\end{equation}
where $t_b$ is measured in days and $a$ and  $b_i$, with $i=1,2$, are calibration constants.

The other one is the Ghirlanda relation, \cite{GGL}.
It connects the collimation-corrected energy, or the energy release of a GRB jet $E_{gamma}$, with the peak energy $E_p$
\begin{equation}\label{eq:noGGL}
 \log{E_{\gamma}} = a + b \log{\frac{E_p}{300 keV}},
\end{equation}
where $a$ and $b$ are two calibration constants.
The $E_{\gamma}$ term depends directly on the break time $t_b$ and the isotropic energy in the following way:
\begin{equation}
 E_{\gamma} = F_{beam} E_{iso}\,.
\end{equation}
where $F_{beam} = 1 - \cos{\theta}$, with $\theta_{jet}$ the jet opening angle defined by \cite{Sari}:
\begin{equation}
 \theta_{jet} = 0.163\left(\frac{t_b}{1 + z}\right)^{3/8}\left(\frac{n_0\eta_{\gamma}}{E_{iso,52}}\right)^{1/8},
\end{equation}
where $E_{iso,52} = E_{iso}/10^{52}$ ergs,  $n_0$ is the
circumburst particle density in 1 cm$^{-3}$, and $\eta_{\gamma}$ the radiative efficiency.

Nevertheless the calibration of these relations has been necessary in order to avoid the circularity problem.
This means that all the relations need to be calibrated for every set of cosmological parameters.
However, currently, there is no a sample of low redshift GRBs, up to z = 0.2-0.3, to allow a cosmology-independent calibration of these relations.
In order to overcome this difficulty, Liang et al., \cite{Liang}, proposed a method in which they calibrated several GRB relations using an interpolation method applied to the Supernova Ia data.
In this way, it becomes possible to build a GRB-Hubble Diagram by computing the luminosity distance for each GRB by
\begin{equation}
\label{lum1}
 d_l = \left(\frac{E_{iso}}{4\pi S_{bolo}'}\right)^{\frac{1}{2}},
\end{equation}
where $E_{iso}$ is the  isotropic energy emitted in the burst and $S_{bolo}'$ is the bolometric fluence corrected to the rest frame of the source in consideration.
The final step consists in calculating the distance modulus $\mu = 25 + 5 \log{d_l}$, and its error, for each GRB.

Our data sample is constituted by 27 GRBs taken from the Schaefer sample of 69 GRB, \cite{Schaefer}.
For each GRB we assume the same value for the radiative efficiency, $\eta_{\gamma} = 0.2$.
Moreover we assume that the error in the determination of the redshift $z$ is negligible, as well as for the same radiative efficiency.
The error bars on the distance modulus are photometric uncertainties only, neglecting any other type of error, e.g. GRB intrinsic variability, so:
\begin{equation}
 \sigma_{\mu} =  \left[\left(2.5 \sigma_{\log{E_{iso}}}\right)^2  + \left(1.086\sigma_{S_{bolo}}/S_{bolo}\right)^2\right]^{\frac{1}{2}}
\end{equation}
with $\sigma_{\log{E_{iso}}}$ and $\sigma_{S_{bolo}}$ obtained from the error propagation applied to Eq.(\ref{eq:noLZ}) and Eq.(\ref{eq:noGGL}).
We note also that the assumptions of a well-known $n_0$ is a strong hypothesis since the goodness of the fit depends also on this parameter.
For this reason we consider the $n_0$ values for each GRB given in the Table 2 of \cite{CI}, also if it lacks, as we said previously, a complete and clear physical basis for the considered relations.

As a final step we simply add a sample of Supernova data to our GRB sample, in order to constraining better the following data fits at low redshifts.
We take in consideration the UNION catalog, compiled by the Supernova Cosmology Project, \cite{SCP}.

\section{DATA FITTING}

So we have obtained a tool capable of testing the various cosmological models up to a redshift z = 7-8.
As a test instrument, we first performed an evaluation of the various cosmographic parameters related to Hubble's Series and then we estimated the trend of the equation of state of the Dark Energy.
The Hubble Series is the Taylor expansion of the Hubble law:
\begin{eqnarray}\label{eq:Visser}
d_l(z) = d_H z \Bigg\{ 1 + {1\over2}\left[1-q_0\right] {z}
\\
-{1\over6}\left[1-q_0-3q_0^2+j_0+ \frac{k \; d_H^2}{a_0^2} \right] z^2
\nonumber
\\
{+}
{1\over24}[
2-2q_0-15q_0^2-15q_0^3+5j_0(1+2q_0)
\nonumber
\\
+ s_0 + \frac{2\; k \; d_H^2 \; (1+3q_0)}{a_0^2}]\; z^3 +
\mathcal{O}(z^4) \Bigg\}
\end{eqnarray}
We stopped at the fourth order in the expansion so that the various terms in the series are related with the cosmographic parameters, defined as follows:
\begin{equation}
q(t) = - {1\over a} \; {d^2 a\over d t^2}  \;\left[ {1\over a} \;
{d a \over  d t}\right]^{-2}\,,
\end{equation}
\begin{equation}
j(t) = + {1\over a} \; {d^3 a \over d t^3}  \; \left[ {1\over a}
\; {d a \over  d t}\right]^{-3}\,,
\end{equation}
\begin{equation}
s(t) = + {1\over a} \; {d^4 a \over d t^4}  \; \left[ {1\over a}
\; {d a \over  d t}\right]^{-4}\,.
\end{equation}
where $t$ represents the cosmic time.
These parameters can be expressed in terms of the cosmological density $1 - \Omega_{\Lambda}$ and the Equation of State (EoS) $w = p/\rho$ of the Dark Energy, \cite{CapozzPRD}, where $p$ and $\rho$ are respectively the pressure and the density of the fluid constituting the Dark Energy.
Using eq.(\ref{eq:Visser}) as theoretical model, we start our data fit in order to estimate the cosmographic parameter $q_0$, $j_0$ and $s_0$.
The results of this fit are given in Table \ref{table:no1}.
\begin{table}
\caption{Results of the fits. LZ is for Liang-Zhang relation, GGL for the Ghirlandaet al. one} 
\label{table:no1} 
\centering 
\begin{tabular}{c c c c} 
\hline\hline 
Fit & $q_0$ & $j_0 + \Omega$ & $s_0$ \\ 
\hline 
LZ & $-0.68 \pm 0.30$ & $0.021 \pm 1.07$ & $3.39 \pm 17.13$  \\
GGL & $-0.78 \pm 0.20$ & $0.62 \pm 0.86$ & $8.32 \pm 12.16$ \\
\hline 
\end{tabular}
\end{table}

An immediate application of these results consists in the estimation of the cosmological density and Dark Energy EoS parameters.
Nevertheless we don't know the exact evolution of the Dark Energy EoS with the redshift, so that we need a parametrization for the function $w$.
For this reason we consider the Chevalier-Polarski-Linder (CPL) parametrization defined as follows
\begin{equation}
 w(z)_{DE} = w_0 + w_a z \left(\frac{1}{1 + z}\right).
\end{equation}

With this function it is possible reconstruct the cosmological evolution of the Dark Energy and, more important, come back to the epoch of the transition acceleration-deceleration of the universe.
Our previous fit results constrain the cosmological parameters to vary in the ranges given in Table \ref{table:no2}.
\begin{table}
\caption{Cosmological density parameters} 
\label{table:no2} 
\centering 
\begin{tabular}{c c c} 
\hline\hline 
Fit & $\Omega_M$ & $\Omega_{\Lambda}$ \\ 
\hline 
LZ & $0.37 \pm 0.31$ & $0.63 \pm 1.13$ \\
GGL & $0.28 \pm 0.30$ & $0.72 \pm 1.09$ \\
\hline 
\end{tabular}
\end{table}
while for the CPL parameters we obtain
\begin{equation}
 w_0 = -0.53 \pm 0.64 \qquad w_a = 0.59 \pm 0.77,
\end{equation}
that within the errors agree with the $\Lambda$CDM concordance model but it doesn't agree with the epoch of the transition acceleration-deceleration : $z_{tr} \approx 10$.

This estimate does not agree very well with the real value of the Dark Energy EoS, since it is estimated that the transition redshift is much earlier than obtained from our simulations.
This is also because the method used here works very well only at $z < 1$.
For this reason we need an analytical formulation of the Hubble diagram valid, in principle, at any redshift, \cite{Izzo}.
Let's write the Friedmann equation in the case of a quasi-flat universe, so with $k \approx 0$:
\begin{equation}\label{eq:no1}
 H^2(z) = H_0^2 \left( 1 + z\right)^{3(w+1)}\,.
\end{equation}
Now the w-parameter indicates the EoS $w = p/\rho$, where $p$ and $\rho$ are the pressure and the matter-energy density of the Universe, respectively.
Using the CPL parametrization for $w$, that in this case is the EoS for a fluid representing the total Universe, then we rewrite the Friedmann equation as follows:
\begin{equation}
 H(z) = H_0 \left[\left(1+z\right)^{\frac{3}{2}(w_0 + w_a + 1)} \exp{\left(\frac{-3w_a z}{2(1+z)}\right)}\right].
\end{equation}
In this way an analytical expression for the luminosity distance, and also for the distance modulus, can be achieved:
\begin{eqnarray}\label{eq:Luca}
 D_l(z) =  \left(\frac{3 w_a}{2}\right)^{-\frac{1+3 w_0+3 w_a}{2}}
\nonumber
\\
\left. \exp{\left(\frac{3 w_a}{2}\right)} \Gamma \left[ \frac{1+ 3 w_0 + 3 w_a}{2}, \frac{3 w_a}{2 (1+\xi)}\right]
\right|_{\xi = 0}^{\xi = z}.
\end{eqnarray}

Substituting  such an expression in the distance modulus, we obtain a model for  data fitting which could work, in principle, at any $z$.
It is important to stress that the obtained expression for the Hubble parameter $H(z)$ is independent of the density parameters, $\Omega_M$ and $\Omega_{\Lambda}$.
Moreover it is also important to emphasize that we use the CPL parameterization not only for the dark energy component, as is done in the previous analysis, but it is used here for the total energy-matter density of the Universe, not just because matter (dark and baryonic) is contributing by a null pressure and the radiation contribute is null, but also because we obtain an exact analytical formulation for the luminosity distance at any redshift $z$.

\begin{figure}
\includegraphics[width=7 cm, height=5 cm]{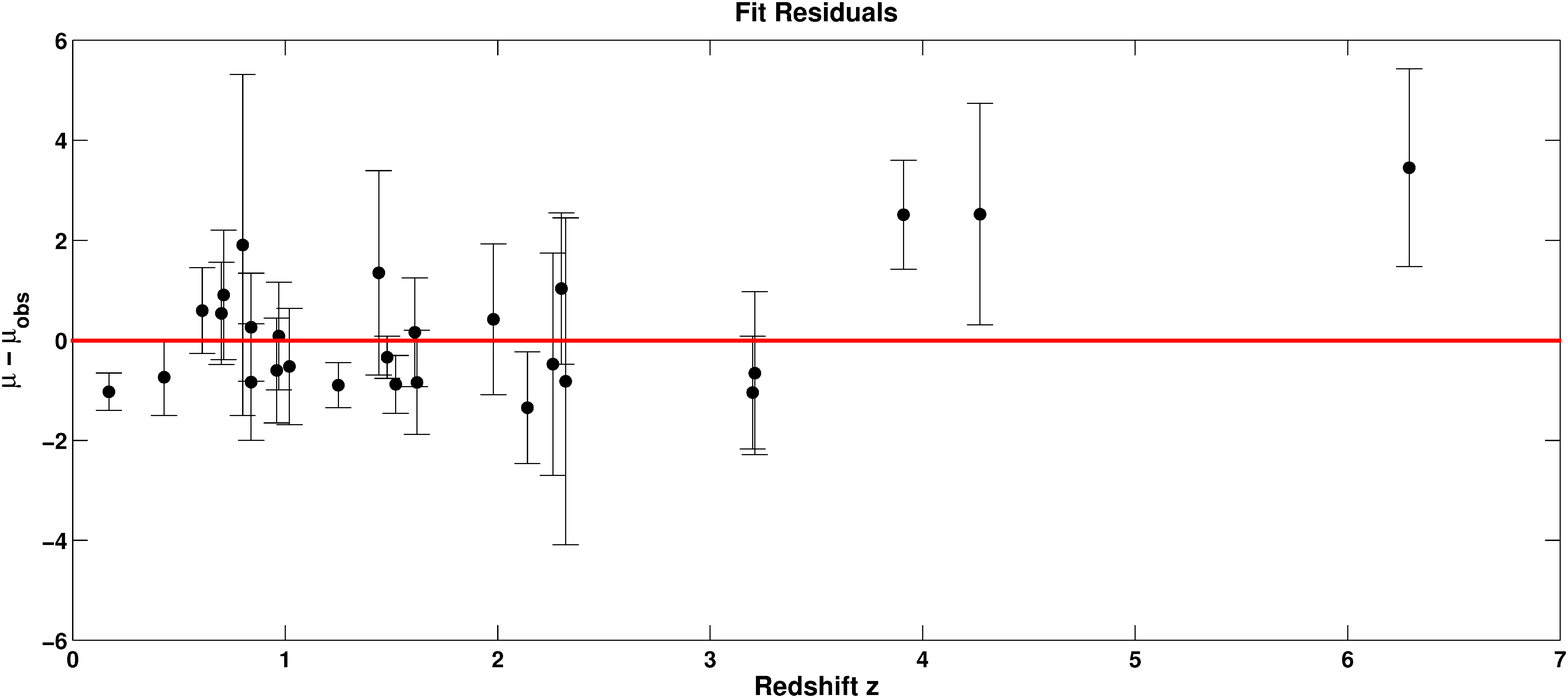}
\caption{Comparison between the best fit of $\mu$ and the
observed distance modulus $\mu_{obs}$ at any redshift. The black
dots are the GRBs data and the red line is the best fit
curve representing the theoretical distance modulus.}
\label{fig:no2}
\end{figure}

With this in mind we start the fit of our combined sample and the results are given in Table \ref{table:no3}.
\begin{table}
\caption{Results of the fits. SNeIa is just for the Supernova Ia data, LZ is for the GRBs data obtained from the Liang-Zhang relation, GGL for the Ghirlanda et al. one. Note the improvement on the w-parameter using the GRBs data in addition to the well-known SNeIa data. corrected for the 3 ``wrong'' GRBs. SNeIa is just for the Supernova Ia data, LZ is for the GRBs data obtained from the Liang-Zhang relation, GGL for the Ghirlanda et al. one.} 
\label{table:no3} 
\centering 
\begin{tabular}{l c c} 
\hline\hline 
Relation & $w_0$ & $w_a$ \\ 
\hline 
 SNeIa & $-0.9097 \pm 0.07$ & $0.755 \pm 0.054$  \\
 LZ & $-1.39 \pm 0.38$ & $1.18 \pm 0.37$  \\
 GGL & $-1.46 \pm 0.38$ & $1.36 \pm 0.32$  \\
 LZ + SNeIa & $-1.15 \pm 0.10$ & $0.93 \pm 0.11$  \\
 GGL + SNeIa & $-1.42 \pm 0.12$ & $1.24 \pm 0.13$  \\
\hline 
\end{tabular}
\end{table}
Immediately we note that the fit results differ if we consider only the Supernova Ia sample from those computed for the sample composed also by the GRBs.
The residual analysis, see Fig. \ref{fig:no2}, gives us the solution to this problem: a smooth trend up to $z \sim 3.5$ in the data distribution can be detected, but beyond this limit we have 3 GRBs that exceed the $3 \sigma$ confidence limit.
It seems like a sort of darkening at high redshift in the GRB luminosity detected.
This fact is fundamental for the goodness of the fit.
Indeed these GRBs represent the most distant object in our Hubble Diagram, so their weight on the fit is very high.
Possible explanations for this anomalous brightness could be various.
In particular there could be some process of absorption of gamma radiation by low energy photons incoming from the cosmic thermal background, \cite{Zdziarski}.
Nevertheless the CPL parametrization could be a bad approximation for the cosmological EoS so that we should consider this problem as due to a uncorrect starting model.

\begin{center}
\begin{figure}
\includegraphics[width=7 cm, height=5 cm]{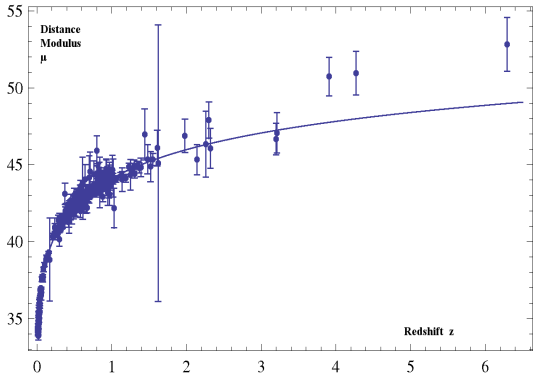}
\caption{Redshift-Distance modulus diagram for the SNeIa+GRB sample. The blue line represents the data fit.}
\label{fig:no1}
\end{figure}
\end{center}

Anyway we reject these 3 GRBs and repeat the analysis without the 3 outliers GRB.
The results of the fits are given in Table \ref{table:no3}.
\begin{table}
\caption{Results of the fits corrected for the 3 ``outlier'' GRBs. SNeIa is for the Supernova Ia data, LZ is for the GRBs data obtained from the Liang-Zhang relation, GGL for the Ghirlanda et al. one.} 
\label{table:no4} 
\centering 
\begin{tabular}{l c c c} 
\hline\hline 
Relation & $w_0$ & $w_a$  \\ 
\hline 
 LZ + SNeIa & $-0.95 \pm 0.01$ & $0.74 \pm 0.01$  \\
 GGL + SNeIa & $-0.865 \pm 0.005$ & $0.66 \pm 0.005$  \\
\hline 
\end{tabular}
\end{table}

\begin{center}
\begin{figure}[ht]
\includegraphics[width=7.5cm, height=5 cm]{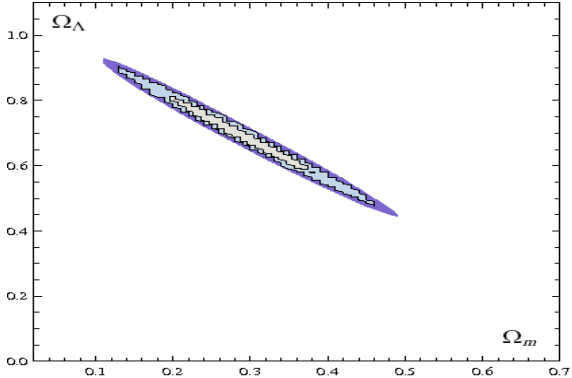}
\caption{68\%, 95\% and 98\% constraints on $\Omega_m$ and $\Omega_{\Lambda}$, obtained from UNION compilation and the GRB sample corrected for the 3 wrong GRBs. }
\label{fig:MCcosmo}
\end{figure}
\end{center}

From a first analysis of the fit results the combined, and corrected, GRB+Sn sample is totally in agree with the concordance model for the present epoch.
This is confirmed also by a Monte Carlo-like procedure for the comparison of the results with the usual likelihood estimator.
In this case we consider, for a sake of completeness, also a small contribute to the total density of the universe due to the curvature density.
The results of this procedure are given in Table \ref{table:no4} and the contour plot, where the boundaries correspond to $1\sigma$, $2\sigma$ and $3\sigma$ confidence levels, are plotted in Fig. (\ref{fig:MCcosmo}).
\begin{table*}[ht]
\caption{Cosmological density parameters, with uncertainties computed to 1$\sigma$ confidence limit, obtained from the MC-like procedure} 
\label{table:no5} 
\centering 
\begin{tabular}{l c c c c} 
\hline\hline 
Sample & $\Omega_m$ & $\Omega_{\Lambda}$ & $\Omega_k$ & $\chi^2$ \\ 
\hline 
 UNION + GRB & $0.26 \pm 0.14$ & $0.73 \pm 0.14$ & $0.01 \pm 0.04$ & $1.032$ \\
 UNION + GRB corrected & $0.25 \pm 0.10$ & $0.74 \pm 0.135 $ & $0.01 \pm 0.035$ & $1.00027$ \\ 
\hline 
\end{tabular}
\end{table*}

\begin{center}
\begin{figure}[ht]
\includegraphics[width=7.5cm, height=5 cm]{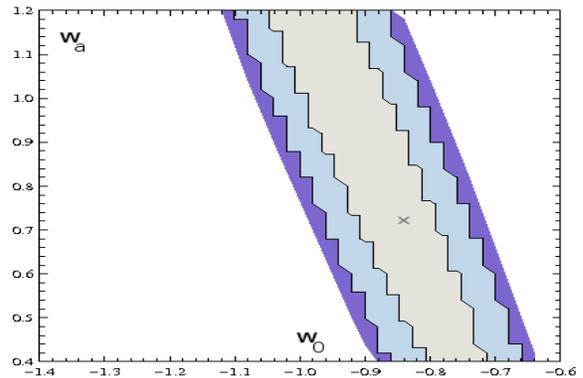}
\caption{68\%, 95\% and 98\% constraints on $w_0$ and $w_a$ obtained from UNION compilation and the GRB sample corrected for the 3 wrong GRBs. The cross represents the best fit value and it is in a good agreement with what found using the theoretical model described in Sect.3.}
\label{fig:MCEoS}
\end{figure}
\end{center}

We have adopted a similar procedure in the case of an EoS evolving with redshift.
The results of this analysis is plotted in Fig. (\ref{fig:MCEoS}) where the best fit value, the cross in the figure, corresponds to the value $w_0 = -0.84 \pm 0.14$ and $w_a = 0.72 \pm 0.06$, in a good agreement with the results obtained using our theoretical model, Eq. \ref{eq:Luca}.
In particular we obtain for the epoch of the transition acceleration-deceleration the value $z_{tr} = 5.17 \pm 0.133$, a result that, if higher than the redshift of the farther GRB used, could be in agreement with current quasar formation scenario.

\section{CONCLUSIONS}

From this analysis we conclude that the corrected GRB+Sn sample agrees fairly woth the $\Lambda$CDM concordance model with a small contribution in terms of curvature density, being this parameter equal to $k = 0.01 \pm 0.04$.
In other words the theoretical model delineated in Eq. \ref{eq:Luca} seems a good approximation of the observed cosmography and agrees very well with the concordance model so that we can argue that GRBs could be good distance indicators at redshift values up to $z = 4$.

However, more robust samples of data are needed and more realistic EoS models, with respect to the simple perfect fluid models, should be taken into account in order to suitably track  redshift at any epoch (see for example \cite{elizalde}).
With the improving of the observations, in particular with the recent launch of new satellites devoted to the GRB surveys, as Fermi-GLAST and AGILE, one should be able to expand the samples of GRBs, possibly by data from objects at higher redshift.

In summary, considering  these preliminary results, it seems that GRBs could be considered as a useful tool to remove degeneration and constrain self-consistent cosmological models. Furthermore the matching with other distance indicators  would improve the consistency of the distance-redshift Hubble diagram by extending it up to redshift $7\, -\, 8$ and over.

\end{document}